\documentclass[aps,pre,showpacs,twocolumn,floats]{revtex4}
\usepackage{amsmath}
\usepackage{graphicx,psfig}
\usepackage{dcolumn}
\usepackage{bm}

\input{tcilatex}

\begin{document}

\title{Saddle index properties, singular topology and its relation to thermodynamic
singularities for a $\phi ^{4}$ mean field model}
\author{D. A. Garanin and R. Schilling}
\affiliation{Institut f\"{u}r Physik,
Johannes-Gutenberg-Universit\"{a}t, D-55099 Mainz, Germany}

\author{A. Scala}
\affiliation{Dipartimento di Fisica, Università di Roma ``La
Sapienza" and Center for Statistical Mechanics and Complexity,
INFM Roma 1, Piazzale Aldo Moro 2, 00185 Roma, Italy }

\date{31 March 2004, revised 29 June 2004}

\begin{abstract}
We investigate the potential energy surface of a $\phi ^{4}$ model with
infinite range interactions. All stationary points can be uniquely
characterized by three real numbers $\alpha _{+},$ $\alpha _{0},$ $\alpha
_{-}$ with $\alpha _{+}+\alpha _{0}+\alpha _{-}=1$, provided that the
interaction strength $\mu $ is smaller than a critical value. The saddle
index $n_{s}$ is equal to $\alpha _{0}$ and its distribution function has a
maximum at $n_{s}^{\max }=1/3$. The density $p(e)$ of stationary points with
energy per particle $e$, as well as the Euler characteristic $\chi (e)$, are
singular at a critical energy $e_{c}(\mu ),$ if the external field $H$ is
zero. However, $e_{c}(\mu )\neq \upsilon _{c}(\mu )$, where $\upsilon
_{c}(\mu )$ is the mean potential energy per particle at the thermodynamic
phase transition point $T_{c}$. This proves that previous claims that the
topological and thermodynamic transition points coincide is not valid, in
general. Both types of singularities disappear for $H\neq 0$. The average
saddle index $\bar{n}_{s}$ as function of $e$ decreases monotonically with $%
e $ and vanishes at the ground state energy, only. In contrast, the saddle
index $n_{s}$ as function of the average energy $\bar{e}(n_{s})$ is given by
$n_{s}(\bar{e})=1+4\bar{e}$ (for $H=0$) that vanishes at $\bar{e}%
=-1/4>\upsilon _{\mathrm{0}}$, the ground state energy.
\end{abstract}

\pacs{05.70.Fh, 61.20.Gy, 64.70.Pf}

\maketitle

\section{Introduction}

Topological features play an important role in several branches of physics.
Examples in condensed matter physics are discussed in Ref.~\cite{1}. Those
examples do not include thermodynamics and phase transitions. That
topological concepts might be relevant for \textit{equilibrium phase
transitions} has already been emphasized long time ago \cite{2} and that
they can be very useful in condensed matter physics has been demonstrated
recently \cite{3,4}.

Usually, equilibrium phase transitions are indicated by a singularity at the
transition temperature $T_{c}$ of thermodynamic quantities, like free
energy, specific heat, etc. One may ask whether other indications for such
phase transitions really exist. This question has been studied by several
groups in recent years. Geometrical entities like the Ricci curvature, and
dynamical ones like Lyapunov exponents were used for a classical planar
Heisenberg model with nearest neighbor interactions and dimension $d=2,3$
\cite{5}, a nearest-neighbor $\phi ^{4}$ model for $d=3$ with O$(n)$%
-symmetry $(n=1,2,4)$ \cite{6} and in $d=1,2$ with O$(1)$-symmetry \cite{7}.
It has been found that both quantities exhibit a singularity at a critical
energy per particle, $e_{c}$, for those dimensions for which an equilibrium
phase transition occurs at $T_{c}>0$. Furthermore, $\upsilon _{c},$ the
internal energy per particle at $T_{c}$, equals $e_{c}$, i.e., the
geometrical and thermodynamic singularity occur at the \textit{same} energy.
It has also been speculated \cite{6,7} that these singularities are related
to qualitative changes in the topology of the potential energy surface (PES)
of those models. That this is true indeed has been proven first for a
\textit{mean-field} XY-model \cite{8}.

One of the most interesting topological quantities is the Euler
characteristic $\chi $ \cite{3,4,9} which is a topological invariant. For
the two-dimensional nearest neighbor $\phi ^{4}$ model \cite{10}, the
mean-field XY \cite{11} and mean-field $k$-trigonometric model \cite{12,13}
it was proven that $\chi (e)$ also becomes singular at $e_{c}=\upsilon _{c}$%
. In addition, it was shown \cite{12,13} that the type of singularity
depends on the order of the phase transition.

$\chi (e)$ is directly related to $M(e,N_{s}),$ the number of stationary
points of the PES of a $N$-particle system with energy $\leq e$ and saddle
index $n_{s}=N_{s}/N$\cite{9},
\begin{equation}
\chi (e)=\sum\limits_{N_{s}=0}^{N}(-1)^{N_{s}}M(e,N_{s}).  \label{eq1}
\end{equation}
$N_{s}$ is the Morse index, i.e., the number of negative eigenvalues of the
corresponding Hessian matrix of a stationary point. $M(e,N_{s})$ is an
exponentially large number in $N\gg 1$:
\begin{equation}
M(e,N_{s})\sim \exp [Ns(e,n_{s})],  \label{eq2}
\end{equation}
where $s(e,n_{s})$ is the configurational entropy per particle. The
corresponding density of states $p(e,n_{s})$ is given by
\begin{equation}
p(e,n_{s})=\frac{\partial }{\partial e}M(e,N_{s})\sim \exp [Ns(e,n_{s})].
\label{eq3}
\end{equation}
The relationship between Eqs. (\ref{eq1}) and (\ref{eq3}) leads to the
assumption that the singularity of $\chi (e)$ may originate from a specific
behavior of $p(e,n_{s})$ for $N\rightarrow \infty $ or that of $s(e,n_{s})$.

The density of states $p(e,n_{s})$ plays an important role in the
investigation of the PES. For instance, one can define the average saddle
index for a fixed energy
\begin{equation}
\bar{n}_{s}(e)=\int_{0}^{1}dn_{s}n_{s}p(e,n_{s})  \label{nsAvrDef}
\end{equation}
and the average energy for a fixed value of $n_{s}$%
\begin{equation}
\bar{e}(n_{s})=\int_{-\infty }^{\infty }de\,\,e\,\,p(e,n_{s}).
\label{eAvrDef}
\end{equation}
In the limit $N\rightarrow \infty $ the averages $\bar{n}_{s}(e)$ and $%
\bar{e}(n_{s})$ are simply the values that maximize the configurational
entropy, i.e., they are solutions of the equations
\begin{equation}
\frac{\partial s}{\partial n_{s}}(e,\bar{n}_{s})=0,\qquad \frac{\partial s}{%
\partial e}(\bar{e},n_{s})=0.  \label{smaxEqs}
\end{equation}

The saddle index properties of a PES have also played an important role in
another respect. Studying \textit{glassy dynamics} and ideal \textit{%
dynamical glass transition} \cite{14} for liquids it has been found
numerically \cite{15,16} that the temperature-dependent average saddle index
practically vanishes at a temperature $T^{\ast }$, which is close to the
mode-coupling glass transition temperature $T_{c}$ \cite{14}. However, this
conclusion should be taken with some care. First of all most saddle points
were quasi saddles (see the discussion in Refs.~\cite{17,18,19}) and second,
plotting $\log \bar{n}_{s}(T)$ (or a related quantity) versus $1/T$ does not
exhibit a quasi singular behavior at $T_{c}$ \cite{20,21}. That $\bar{n}%
_{s}(T)$ vanishes at $T=0$, only, has been proven for the $k$-trigonometric
model \cite{13}. This indicates that $\bar{n}_{s}(T)$ for systems with \emph{%
self-generated} disorder behaves differently than for systems where the
disorder is \emph{quenched}. For the latter it has been proven that $\bar{n}%
_{s}(T)$ vanishes at the dynamical transition point, at least for mean field
like models \cite{CGP}.

A numerical determination of the \textit{true} saddles of a binary liquid
with particle number $N\leq 13$ gives evidence that $\bar{n}_{s}(e)$
vanishes at an energy $e^{\ast }$, which still depends on $N$ \cite{17}.
This evidence holds for both cases where the average $\bar{n}_{s}$ is
plotted versus $e$ and $n_{s}$ is plotted versus $\bar{e}$ \cite{17}.
Whether this vanishing at $e^{\ast }$ is spurious or not is not known. It is
obvious why vanishing of $\bar{n}_{s}$ at $T^{\ast }$ or $e^{\ast }$ may be
relevant. In that case, the system is mostly close to local minima $%
(n_{s}=0) $ for $T<T^{\ast }$ or $e<e^{\ast }$ and the dynamics is dominated
by activated processes, in contrast to $T>T^{\ast }$ or $e>e^{\ast }$, where
the particles dynamics is more flow-like. Hence, vanishing of $\bar{n}%
_{s}(e) $ may indicate a qualitative change in the dynamics \cite{15,16}.

There is another result presented in Ref.~\cite{17} which concerns the
distribution function of the saddle index $p(n_{s})$:
\begin{equation}
p(n_{s})=\int\limits_{-\infty }^{\infty }dep(e,n_{s}).  \label{eq4}
\end{equation}
It is found that $p(n_{s})$ is a Gaussian with a maximum at $n_{s}^{\mathrm{%
max}}\approx 1/3$. Although the physical relevance of this result is not
clear, it seems to be an interesting property of the topology of the PES.

We hope that the exposition above has made obvious the role of topological
features for both the thermodynamic and dynamic behavior. It is the main
purpose of our paper to analytically investigate for a \textit{mean-field} $%
\phi ^{4}$ model the existence of a singularity in the topology of its PES
and the relation to a thermodynamic singularity and to calculate the saddle
index properties discussed above.

The outline is as follows. The mean-field $\phi ^{4}$ model and its basic
properties will be discussed in the Sec. \ref{sec-2}. In Sec. \ref{sec-3} we
will investigate the topological properties of the model. In particular, we
will prove that the claim that topological and thermodynamic transition
points coincide is not correct, in general. The final section contains
discussion where we explain the origin of this discrepancy. Some more
technical details are given in the Appendix.

\section{Mean-field $\protect\phi ^{4}$ model}

\label{sec-2}

Let $x_{n}$ be a scalar displacement of a particle from a lattice site $n$.
We consider the following potential energy
\begin{equation}
V(\mathbf{x,}H)=\sum\limits_{n=1}^{N}V_{0}(x_{n},H)-\frac{\mu }{2N}\left(
\sum\limits_{n=1}^{N}x_{n}\right) ^{2}  \label{eq5}
\end{equation}
depending on the $N$-particle configuration $\mathbf{x}=(x_{1},\ldots
,x_{N}) $. $V_{0}(x,H)$ is an asymmetric on-site potential:
\begin{equation}
V_{0}(x,H)=-xH-\frac{1}{2}x^{2}+\frac{1}{4}x^{4}  \label{eq6}
\end{equation}
that becomes symmetric for $H=0.$ The final term in Eq. (\ref{eq5})
represents the harmonic interaction between \textit{all} particles with a
coupling parameter $\mu \geq 0$. The reader should note that the potential
energy and the displacement can always be scaled such that $V_{0}(x,H)$ has
the $x$-dependence given in Eq. (\ref{eq6}). This type of models was used to
describe structural phase transitions \cite{22}. In contrast to the
mean-field models studied in Refs. \cite{8,11,12,13} there is a nontrivial
coupling constant $\mu $, which can \textit{not} be put to one by an
appropriate scaling of the temperature.

Some thermodynamic properties of the model described by Eqs. (\ref{eq5}) and
(\ref{eq6}) as well as some features of its PES were already investigated
\cite{23}. Let us recall these results and start with the thermodynamic
behavior. Due to the infinite-range interaction the mean-field approximation
becomes \textit{exact} for $N\rightarrow \infty $. This leads to the
self-consistency equation for the order parameter $\left\langle
x\right\rangle =\langle x_{n}\rangle \left( T,H\right) $:
\begin{equation}
\left\langle x\right\rangle =\frac{1}{\mathcal{Z}}\int\limits_{-\infty
}^{\infty }dxx\exp \left[ -\frac{V_{0}(x,H)-\mu \left\langle x\right\rangle x%
}{T}\right] ,  \label{eq7}
\end{equation}
where $\mathcal{Z}$ is the corresponding partition function and $\beta =1/T.$
Of course, a phase transition (second order) occurs at some $T_{c}$ for $H=0$%
, only. $T_{c}$ follows from:
\begin{equation}
T_{c}=\mu \frac{\int\nolimits_{-\infty }^{\infty }dxx^{2}\exp \left[
-V_{0}(x,0)/T_{c}\right] }{\int\nolimits_{-\infty }^{\infty }dx\exp \left[
-V_{0}(x,0)/T_{c}\right] }.  \label{eq9}
\end{equation}
For $0<\mu \ll 1$ one finds:
\begin{equation}
T_{c}(\mu )=\mu +{\mathcal{O}}(\mu ^{2})  \label{eq10}
\end{equation}
which yields for the average potential energy per particle $\bar{\upsilon}%
(T)=\lim\limits_{N\rightarrow \infty }N^{-1}\langle V(\mathbf{x})\rangle
\left( T,H=0\right) $ at $T_{c}$
\begin{equation}
\upsilon _{c}=\upsilon (T_{c})=-\frac{1}{4}(1-2\mu )+{\mathcal{O}}(\mu ^{2})
\label{eq11}
\end{equation}
which is always larger than the minimum value $-1/4$ of $V_{0}(x,0)$.

\begin{figure}
\unitlength1cm
\begin{picture}(11,6)
\centerline{\psfig{file=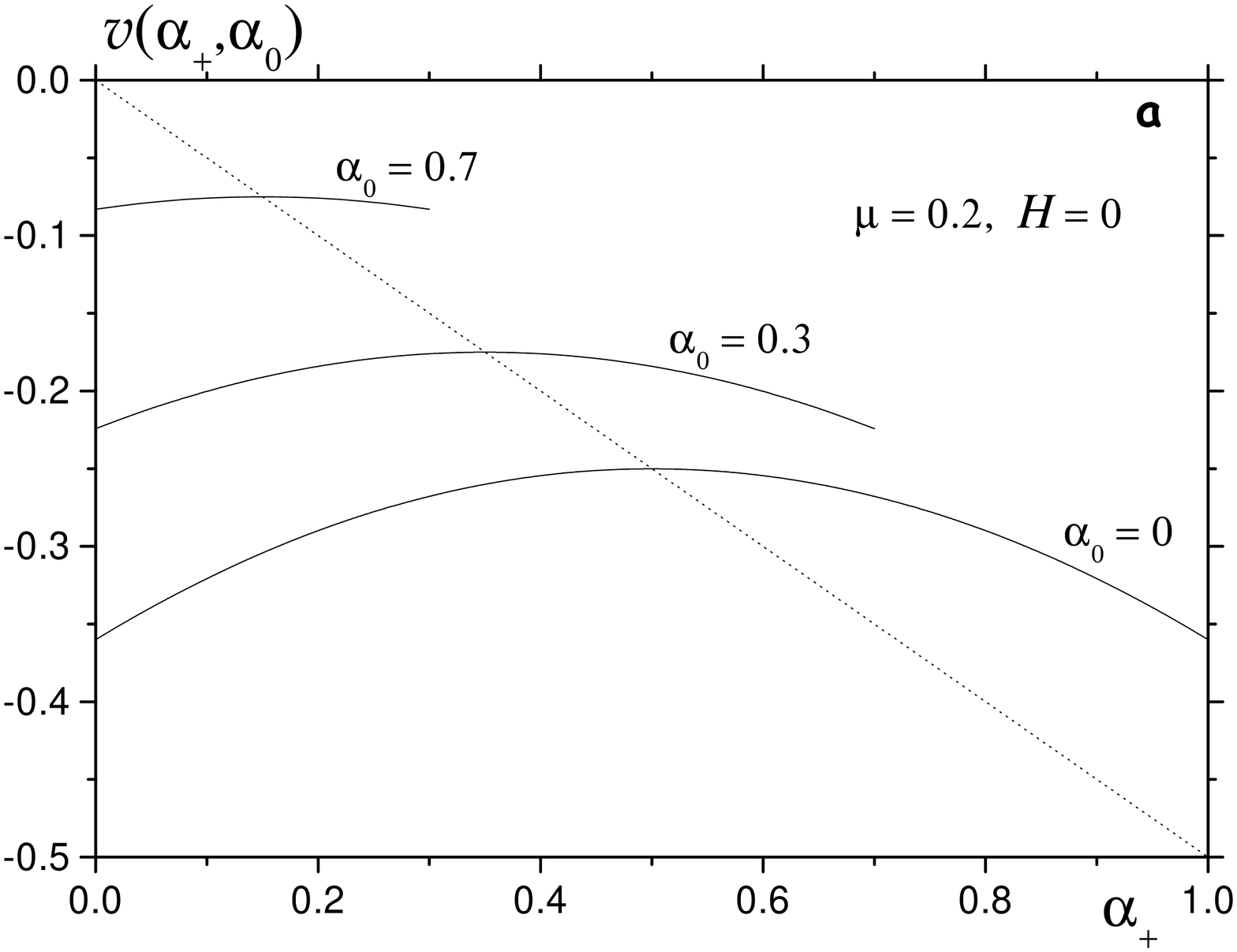,angle=-90,width=9cm}}
\end{picture}
\unitlength1cm
\begin{picture}(11,6)
\centerline{\psfig{file=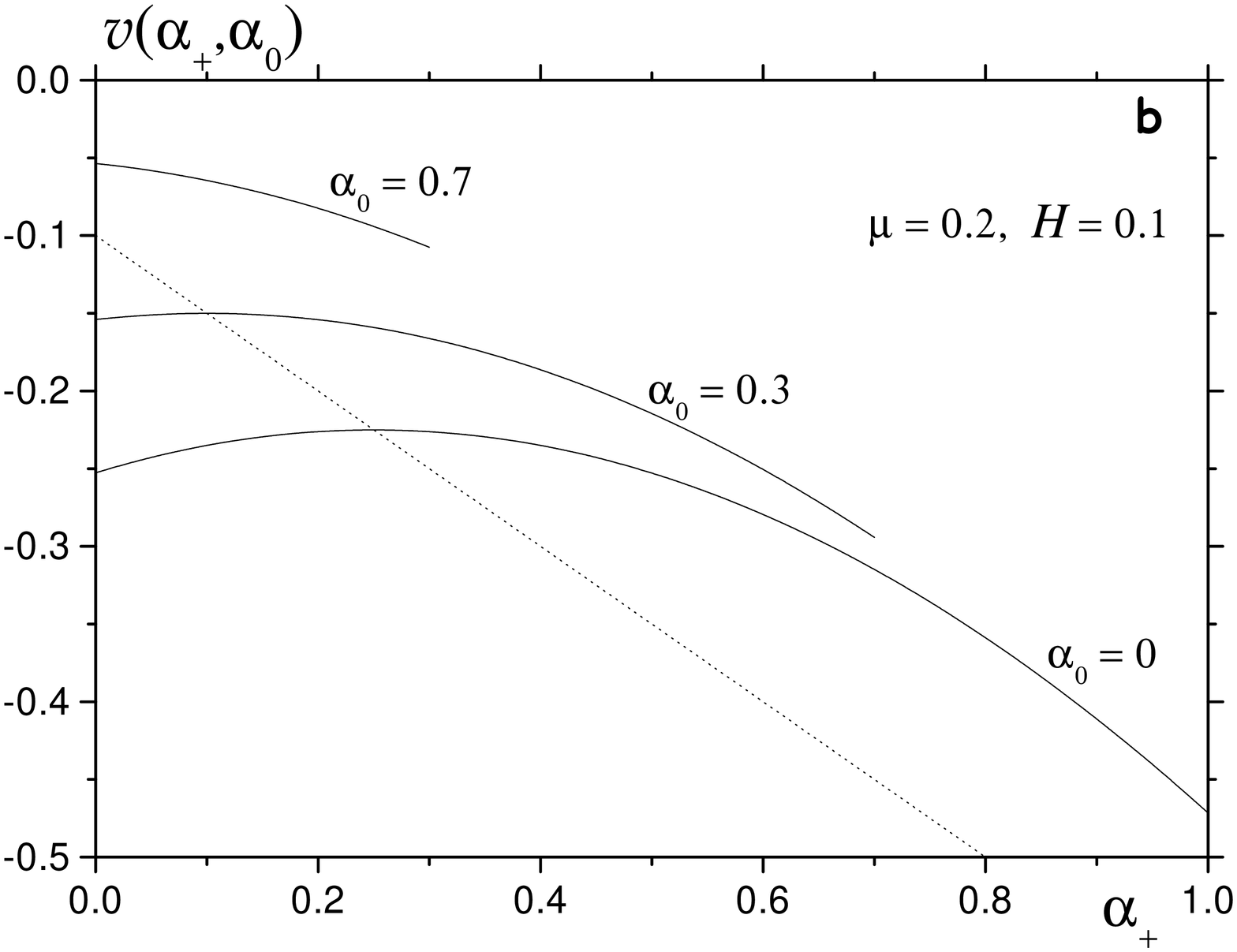,angle=-90,width=9cm}}
\end{picture}
\caption{\label{fig1}
$a$ -- $\alpha_+$-dependence of
$\upsilon(\alpha_+, \alpha_0)$ for $\mu=0.2$ and $H=0$;
$b$ --  Same for $H=0.1$. Dotted lines locate the maxima.
Note the concavity of $\upsilon(\alpha_+, \alpha_0)$ in $\alpha_+$ for $H\geq 0$ and its symmetry
with respect to the maximum position  $\alpha_+^{\max}=(1-\alpha_0)/2$ for $H=0$.}
\end{figure}

Let us now turn to the stationary points as discussed in Ref. \cite{23}.
With the ``internal'' field
\begin{equation}
H_{\mathrm{int}}=\frac{\mu }{N}\,\sum\limits_{n=1}^{N}x_{n}  \label{eq12}
\end{equation}
the solution of $\partial V/\partial x_{n}$ reduces to that of
\begin{equation}
x^{3}-x-H_{\mathrm{eff}}=0  \label{eq13}
\end{equation}
with the effective field \cite{24}
\begin{equation}
H_{\mathrm{eff}}=H+H_{\mathrm{int}}.  \label{eq14}
\end{equation}
For $|H_{\mathrm{eff}}|<H_{c}=2/(3\sqrt{3})$ there are three real roots of
Eq.~(\ref{eq13}) which will be denoted by $x_{\sigma }(H_{\mathrm{eff}})$, $%
\sigma =+,0,-$. It is $x_{+}>x_{0}>x_{-}$. A stationary point of $V$ is
characterized by $N_{\sigma }$, the number of $x_{n}$ in $\mathbf{x}$ which
are equal to $x_{\sigma }(H_{\mathrm{eff}})$. Permuting the particle indices
yields stationary points with the same potential energy. Since $\{x_{n}\}$
are displacements and not positions of particles in a liquid, these
permutations should be counted as \textit{different} stationary points.
Hence there are:
\begin{equation}
P(N_{+},N_{0})=\frac{N!}{N_{+}!N_{0}!N_{-}!}  \label{eq15}
\end{equation}
stationary points of class $(N_{+},N_{0},N_{-})$, where
\begin{equation}
\sum\limits_{\sigma }\,N_{\sigma }=N,  \label{eq16}
\end{equation}
i.e., $N_{-}=N-N_{+}-N_{0}.$ The characterization of all stationary points
by $N_{+},N_{0}$ and $N_{-}$ or equivalently by
\begin{equation}
\alpha _{\sigma }=\frac{N_{\sigma }}{N},\quad \sum\limits_{\sigma }\alpha
_{\sigma }=1  \label{eq17}
\end{equation}
proves (see, e.g.,~Sec. 3) to be extremely useful. Having specified $\alpha
_{\sigma }$ we can determine $H_{\mathrm{eff}}(\alpha _{+},\alpha _{0})$
from Eqs.~(\ref{eq12}) and~(\ref{eq14}):
\begin{equation}
H_{\mathrm{eff}}=H+\mu \,\sum\limits_{\sigma }\alpha _{\sigma }x_{\sigma
}(H_{\mathrm{eff}}).  \label{eq18}
\end{equation}
[We omit the arguments $H$ and $\mu $ for brevity and we take into account
that $\alpha _{-}$ can be expressed by $\alpha _{+},\alpha _{0}$, due to
Eq.~(\ref{eq17}). Also we will mostly drop the arguments $(\alpha
_{+},\alpha _{0})$ of $H_{\mathrm{eff}}$]. Finally the stationary points of
class $(N_{+},N_{0},N_{-})$ are given by:

\begin{equation*}
\mathbf{x}(H_{\mathrm{eff}})=(
\begin{array}[t]{c}
\underbrace{x_{+}(H_{\mathrm{eff}}),\cdots } \\
N_{+}\text{ times}
\end{array}
;
\begin{array}[t]{c}
\underbrace{x_{0}(H_{\mathrm{eff}}),\cdots } \\
N_{0}\text{ times}
\end{array}
;
\begin{array}[t]{c}
\underbrace{x_{-}(H_{\mathrm{eff}}),\cdots } \\
N_{-}\text{ times}
\end{array}
).
\end{equation*}
and its permutations. Similarly for the potential energy per particle $%
\upsilon (\alpha _{+},\alpha _{0})$ follows:
\begin{equation}
\upsilon (\alpha _{+},\alpha _{0})=\sum\limits_{\sigma }\alpha _{\sigma
}V_{0}(x_{\sigma }(H_{\mathrm{eff}}),H)-\frac{1}{2\mu }\left[ H_{\mathrm{eff}%
}-H\right] ^{2}.  \label{vgenaral}
\end{equation}
Fig.~\ref{fig1} presents $\upsilon (\alpha _{+},\alpha _{0})$ as a function
of $\alpha _{+}$ for different values of the parameters.

%

Eq. (\ref{vgenaral}) holds for $|H_{\mathrm{eff}}|\leq H_{c}$. One can
easily prove that the latter is guaranteed for arbitrary $\{\alpha _{\sigma
}\}$ with $\sum\limits_{\sigma }\alpha _{\sigma }=1$ if
\begin{equation}
-H_{c}(1-3\mu )\leq H\leq (1-3\mu )H_{c}.  \label{eq22}
\end{equation}
This implies the \textit{necessary} condition $\mu <1/3$ (cf.~also Ref.~\cite
{23}). $\mu =1/3$ is a critical value for the coupling parameter at which
the PES changes qualitatively. If inequality Eq.~(\ref{eq22}) is violated,
the stationary points are \textit{not} uniquely classified by $\{\alpha
_{\sigma }\}$ \cite{25}. Therefore we will restrict ourselves to $\mu \leq
1/3$, in the following. What remains to be done is to determine the saddle
index $n_{s}$ of these stationary points. The answer is simple \cite{23},
because
\begin{equation}
n_{s}\equiv \alpha _{0}.  \label{nsalpha0}
\end{equation}
This has been proven for $H=0$ in Ref. \cite{23} by determination of the
number $N_{s}$ of negative eigenvalues of the Hessian at stationary
configurations with $N_{0}$ fixed. The result is that $N_{s}=N_{0}$, which
implies Eq.~(\ref{nsalpha0}). Validity of Eq.~(\ref{nsalpha0}) can also be
seen as follows. If $\mu =0$ and $H\geq 0$, the stationary points of $V$ are
given by those of $V_{0}$ and the sign of the eigenvalues of the Hessian
equals the sign of $V_{0}^{^{\prime \prime }}$. Since $N_{0}$ is the number
of particles with position $x_{0}(H)$, which is on the concave part of $%
V_{0} $, i.e. $V_{0}^{^{\prime \prime }}(x_{0}(H))<0$, it is $N_{0}=N_{s}$. $%
N_{s}$ is a topological invariant for $0\leq \mu <1/3$. Consequently Eq.~(%
\ref{nsalpha0}) remains true for all $\mu <1/3$ and all $H$ obeying Eq.~(\ref
{eq22}).

\section{Statistics of stationary points}

\label{sec-3}

\subsection{ Saddle index distribution}

We have shown in Sec. \ref{sec-2} that all stationary states of the
mean-field $\phi ^{4}$ model are characterized by $\{\alpha _{\sigma }\}$,
provided inequality Eq.~(\ref{eq22}) holds. $\alpha _{0}=n_{s}$ is the
saddle index. Consequently there are in total $3^{N}$ stationary points,
from which $2^{N}$ are local minima. The saddle index distribution follows
from Eq.~(\ref{eq15}):
\begin{equation}
p(n_{s})=3^{-N}\,\sum\limits_{N_{+}=0}^{N-N_{s}}P(N_{+},N_{s})=\frac{%
3^{-N}2^{N-N_{s}}N!}{N_{s}!(N-N_{s})!}.  \label{eq28}
\end{equation}
Using the Stirling formula for $N\gg 1$ one obtains
\begin{equation*}
p(n_{s})\sim \exp \left[ Ns(n_{s})\right] ,
\end{equation*}
where the configurational entropy is given by
\begin{equation}
s(n_{s})=-\left[ n_{s}\ln n_{s}+(1-n_{s})\ln \frac{1-n_{s}}{2}\right] .
\label{eq29}
\end{equation}
The maximum of $p(n_{s})$ is at $n_{s}^{\mathrm{max}}=1/3$ which is obvious
since $P(N_{+},N_{0})$ has a maximum at $N_{+}=N_{0}=N_{-}\equiv N/3$. It is
interesting that this result based on the double-well character of the local
potential, coincides with the numerical finding for binary Lennard-Jones
clusters \cite{17}. We stress that the validity of $n_{s}^{\mathrm{max}}=1/3$
is more general. Suppose there are no interactions. Then the $N$-particle
problem separates into that of $N$ independent particles in double wells,
for which $P(N_{+},N_{0})$ is still given by Eq.~(\ref{eq15}). Turning on an
arbitrary interaction will not destroy the one-to-one correspondence between
stationary points and $(N_{+},N_{-},N_{0})$, up to a critical interaction
strength. At this critical strength, e.g.,~exponentially many metastable
configurations may become unstable. Accordingly, $n_{s}^{\mathrm{max}}=1/3$
holds up to that critical coupling, i.e.,~it is a topological invariant.

\subsection{ Calculation of the density of states $p(e,n_{s})$}

The density of stationary points of a PES with energy $e$ and saddle index $%
n_{s}$ that was mentioned in the Introduction is defined by
\begin{equation}
p(e,n_{s})=\sum_{N_{+}=0}^{N-N_{0}}P(N_{+},N_{0})\delta \left( \upsilon
(\alpha _{+},\alpha _{0})-e\right) ,  \label{peDef}
\end{equation}
where $P(N_{+},N_{0})$ is given by Eq. (\ref{eq15}) and
\begin{equation*}
N_{0}=\alpha _{0}N=n_{s}N,\qquad N_{+}=\alpha _{+}N.
\end{equation*}
The density $p(e)$ of stationary points with energy $e$ follows from
\begin{equation}
p(e)=\sum_{N_{0}=0}^{N}p(e,n_{s}),  \label{pensDef}
\end{equation}
Neglecting the irrelevant prefactor one can immediately write ($\alpha
_{0}=n_{s}$)
\begin{equation}
p(e,n_{s})\sim P(N_{+}(e,\alpha _{0}),N_{0}),  \label{pensP}
\end{equation}
where $N_{+}(e,\alpha _{0})=\alpha _{+}(e,\alpha _{0})N$ follows from the
solution of the equation
\begin{equation}
\upsilon (\alpha _{+},\alpha _{0})=e.  \label{eq32}
\end{equation}
This equation has two solutions $\alpha _{+}^{\pm }(\alpha _{0},e)$ which
are derived and discussed in Appendix B.

\begin{figure}
\unitlength1cm
\begin{picture}(11,6)
\centerline{\psfig{file=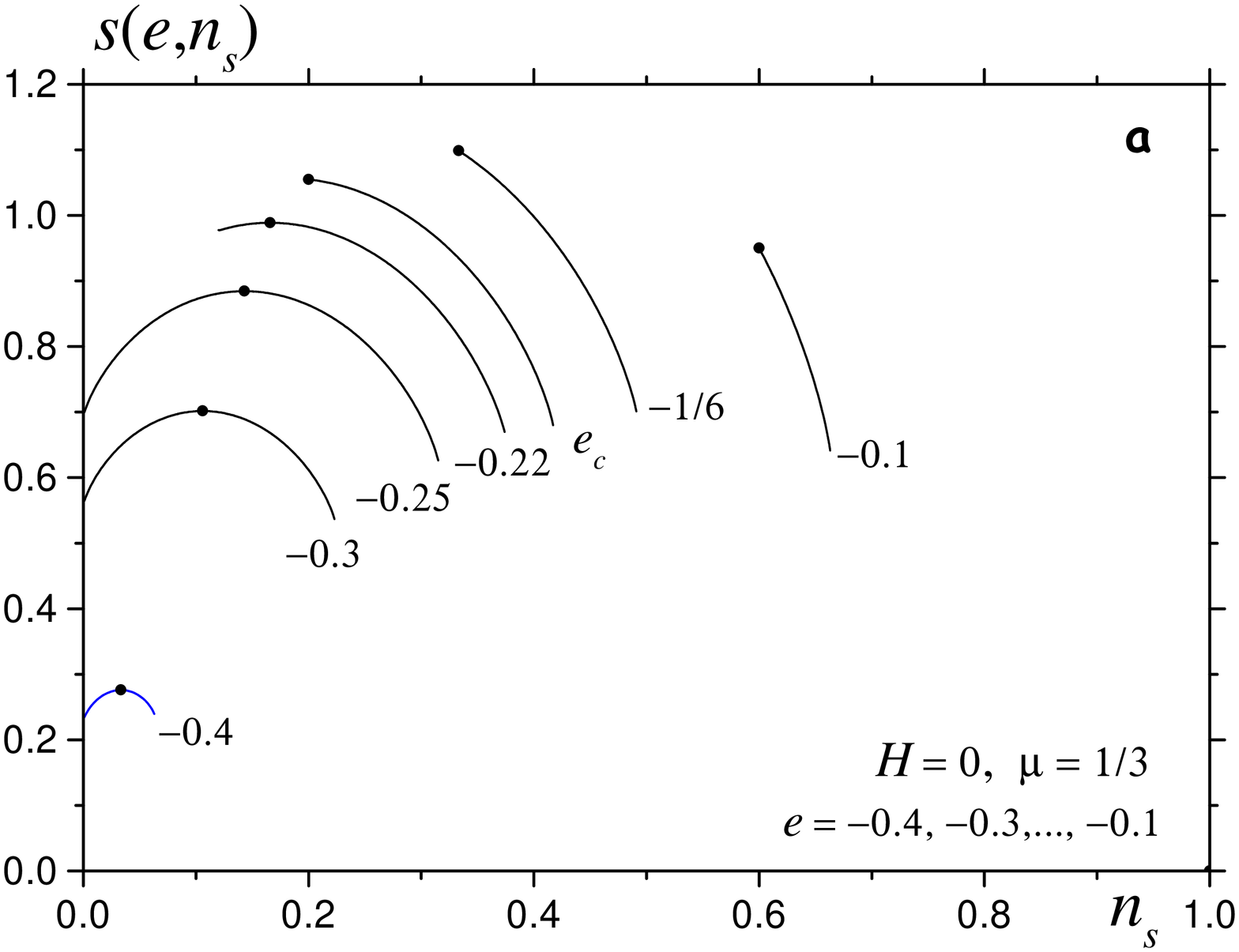,angle=-90,width=9cm}}
\end{picture}
\begin{picture}(11,6)
\centerline{\psfig{file=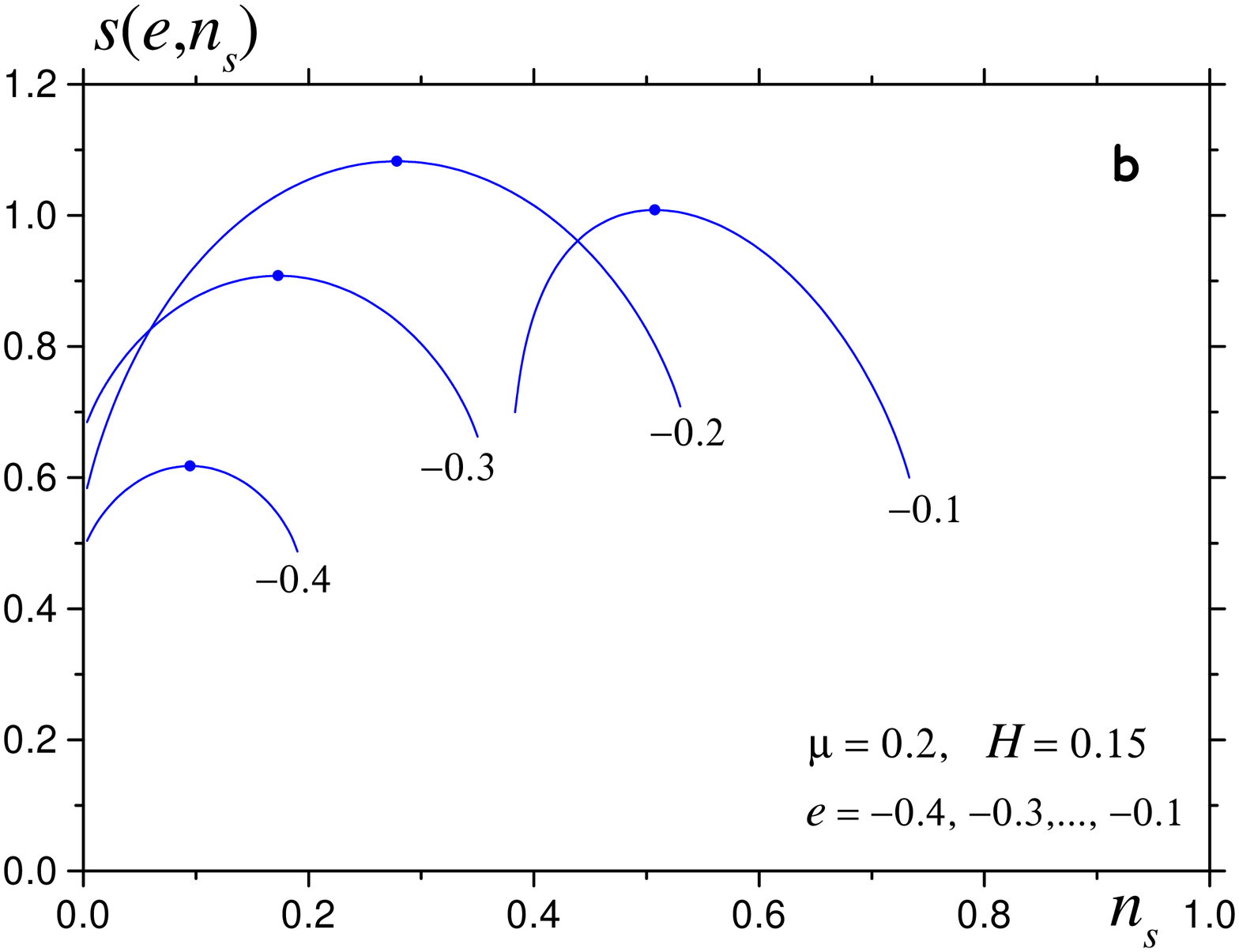,angle=-90,width=9cm}}
\end{picture}
\caption{\label{fig2} $a$ -- Configurational entropy $s(e,n_s)$ versus $n_s$ for different energies and zero field;
$b$ --  same for nonzero  field.}
\end{figure}

With the use of the Stirling formula $P(N_{+},N_{0})$ simplifies to $%
P(N_{+},N_{0})\sim \exp \left[ Ns\left( e,\alpha _{0}\right) \right] ,$ thus
one obtains Eq. (\ref{eq3}) with the configurational entropy given by
\begin{equation}
s\left( e,\alpha _{0}\right) =-\alpha _{0}\ln \alpha _{0}-\alpha
_{+}^{(+)}\ln \alpha _{+}^{(+)}-\alpha _{-}^{(+)}\ln \alpha _{-}^{(+)},
\label{sealpha0}
\end{equation}
where $\alpha _{-}^{(+)}=1-\alpha _{0}-\alpha _{+}^{(+)}$ and we have taken
the (+) branch of Eq.(\ref{alphaplusEAppr}) that makes the dominant
contribution into $P(N_{+},N_{0})$ for $H>0.$ For $H=0$ one can also
restrict oneself to $\alpha _{+}^{(+)},$ due to the symmetry. Note that $%
p(e,n_{s})$ is nonzero and given by Eq. (\ref{eq3}) only in the energy
window
\begin{equation}
\upsilon _{\mathrm{\min }}(\alpha _{0})\leq e\leq \upsilon _{\mathrm{max}%
}(\alpha _{0}),  \label{ewindow}
\end{equation}
where $\upsilon _{\mathrm{\min }}(\alpha _{0})=\upsilon (1-\alpha
_{0},\alpha _{0})$ (see Fig. \ref{fig1}) and $\upsilon _{\mathrm{max}%
}(\alpha _{0})$ is given by Eq. (\ref{eq27}), otherwise $p(e,n_{s})=0.$
Alternatively one can say that Eq. (\ref{eq3}) is valid in the window of
saddle indices
\begin{equation}
\alpha _{0}^{(\min )}(e)\leq \alpha _{0}\leq \alpha _{0}^{(\max )}(e),
\label{alpha0window}
\end{equation}
where the boundary values satisfy $\upsilon _{\mathrm{\min }}(\alpha
_{0}^{(\max )})=e$ and $\upsilon _{\mathrm{max}}(\alpha _{0}^{(\min )})=e.$
From Eq. (\ref{eq27}) one finds
\begin{equation}
\alpha _{0}^{(\min )}(e)=1+4e-2H^{2}/\mu ,  \label{alpha0min}
\end{equation}
whereas $\alpha _{0}^{(\max )}(e)$ can be found with the help of Eq. (\ref
{vgenaral}) or, approximately, with the help of Eq. (\ref{EnergyParabola}).
The dependence of $s(e,n_{s})$ on $n_{s}$ is shown in Fig. \ref{fig2} for
zero and nonzero field.

%

Let us discuss the main features of $s(e,n_{s})$ presented in Fig. \ref{fig2}%
. For a more detailed analytical discussion of $s(e,n_{s})$ the interested
reader is referred to Appendix B. We begin with $H=0$ (see Fig. \ref{fig2}%
a). The maximum of $s(e,n_{s})$ with respect to $n_{s}$ is denoted by $\bar{n%
}_{s}(e)$. Because of the relation between $p(e,n_{s})$ and $s(e,n_{s})$
given by Eq.~(\ref{eq3}) it is obvious that for $N\rightarrow \infty $ the
maximum position $\bar{n}_{s}(e)$ is identical to the averaged saddle index $%
\bar{n}_{s}(e)$ given by Eq.~(\ref{nsAvrDef}). In Appendix B the existence
of a \textit{critical} energy $e_{c}(\mu )$ is proven. $s(e,n_{s})$ as
function of $n_{s}$ has a maximum within the domain of $n_{s}$ for $%
e<e_{c}(\mu )$ and a maximum at the left border of its domain for $e\geq
e_{c}(\mu )$. This implies that the slope $\partial s(e,n_{s})/\partial n_{s}
$ at $n_{s}(e)$ is continuous in $e$, but not differentiable, i.e., the
``curvature'' $\partial ^{2}s(e,n_{s})/\partial n_{s}^{2}$ is discontinuous
in $e$ at $e=e_{c}(\mu )$. This is the origin of the topological
singularity, discussed below. Fig. 3a presents $\bar{n}_{s}(e)$ and reveals
the singularity at $e_{c}(\mu )$. Note that $\bar{n}_{s}(e)$ contains a
branch that is independent of the interaction [see Eqs. (\ref{nsavrResFinaö}%
) and (\ref{nsmax})]. For $e$ very close to the ground state energy $%
v_{0}(\mu )$ [cf. (\ref{eq26})] one obtains the power-law behavior (see
Appendix B):
\begin{equation}
\bar{n}_{s}(e)\sim \lbrack e-v_{0}(\mu )]^{\delta (\mu )}  \label{34}
\end{equation}
with $\delta (\mu )>1$, if $\mu $ is small enough. Note that $\bar{n}%
_{s}(e)\rightarrow 0$ for $e\rightarrow v_{0}(\mu )$. The averaged saddle
index $\bar{n}_{s}(e)$ takes the critical value $n_{s}^{(c)}(\mu )=\bar{n}%
_{s}(e_{c}(\mu )$ shown in Fig. \ref{fig3}. This figure also includes the
asymptotic result of Eq. (\ref{critParAnal}) for $\mu \rightarrow 0$.

\begin{figure}
\unitlength1cm
\begin{picture}(11,6)
\centerline{\psfig{file=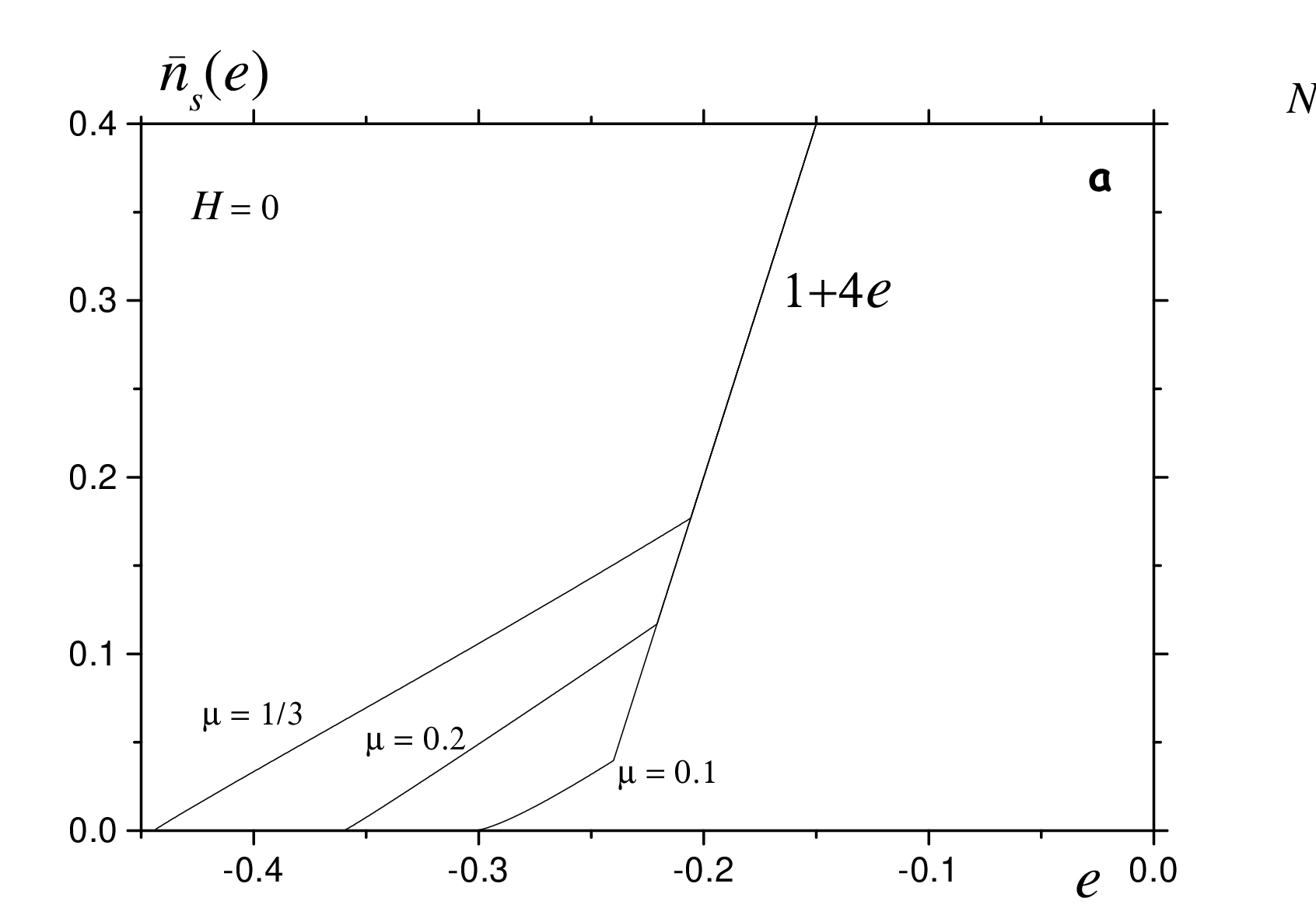,angle=-90,width=9cm}}
\end{picture}
\begin{picture}(11,6)
\centerline{\psfig{file=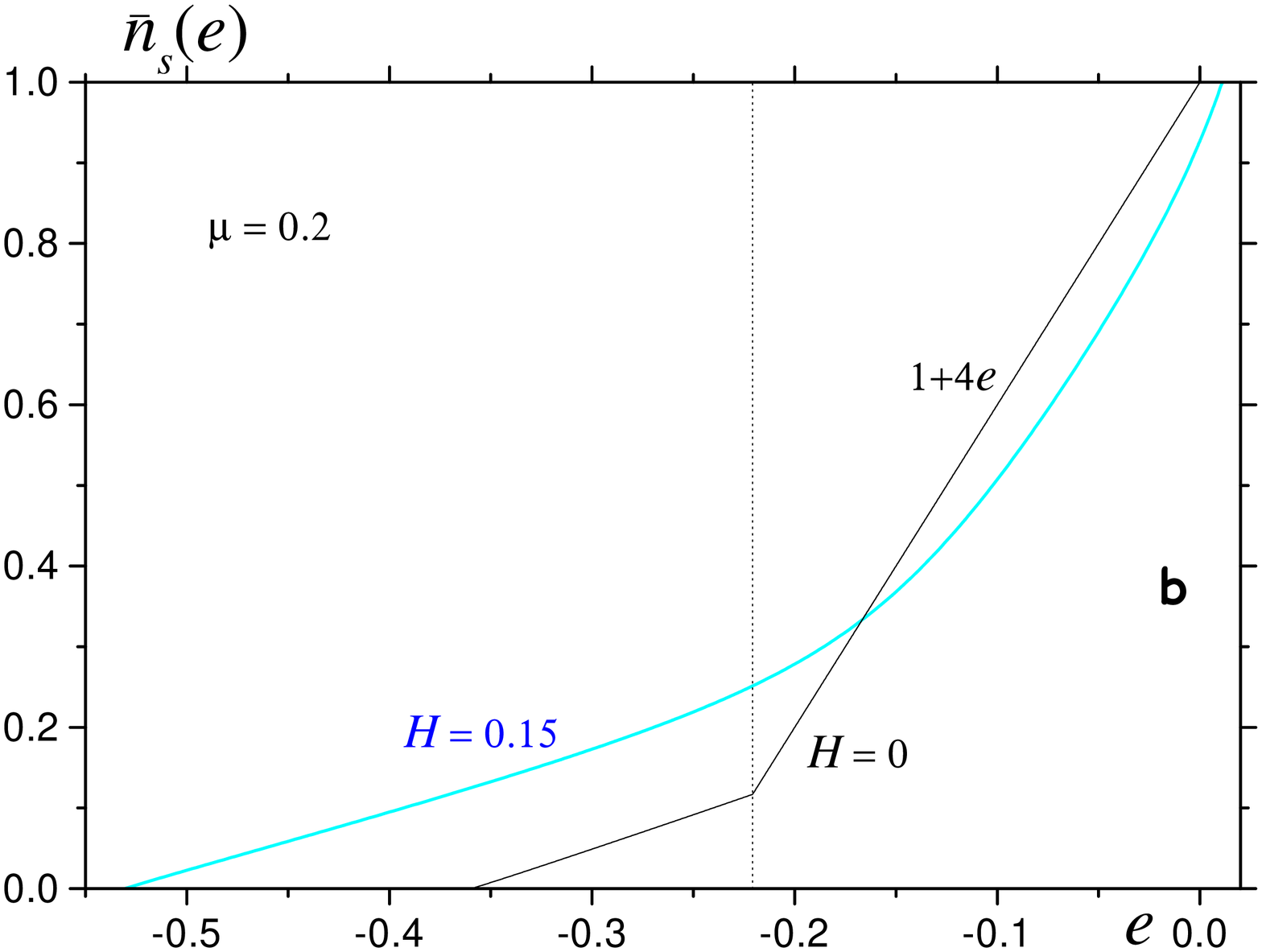,angle=-90,width=9cm}}
\end{picture}
\caption{\label{fig4} $a$ -- Averaged saddle index $\bar n_s(e)$
versus $e$ for $\mu=0.1, 0.2, 1/3$ and $H=0$.;
$b$ -- Same for $\mu=0.2$ and $H=0$ and $H=0.15$}
\end{figure}%
%

Instead of fixing $e,$ one can also determine the maximum position $\bar{e}%
(n_{s})$ of $s(e,n_{s})$ for given $n_{s}$. $\bar{e}(n_{s})$ is the averaged
saddle energy [cf. Eq.~(\ref{eAvrDef})] as function of $n_{s}$. It is easy
to get $\bar{e}(n_{s})$, since the number of stationary configurations is
maximal for $\alpha _{+}=\alpha _{-}=(1-n_{s})/2$. This yields for the
effective field $H_{\text{eff}}(\alpha _{+}=(1-n_{s})/2,\alpha
_{0}=n_{s})\equiv 0$ which implies $x_{\pm }(\alpha _{+}=(1-n_{s})/2,\alpha
_{0}=n_{s})=\pm 1$ and $x_{0}(\alpha _{+}=(1-n_{s})/2,\alpha _{0}=n_{s})=0$
and this in turn leads to

\begin{equation}
\bar{e}(n_{s})\equiv v(\alpha _{+}=(1-n_{s})/2,\alpha
_{0}=n_{s})=-(1-n_{s})/4.  \label{35}
\end{equation}

The reader should note that (i) the inverse function $n_{s}(\bar{e})$ (see
Fig. \ref{fig5}) of $\bar{e}(n_{s})$ turns to zero at $\bar{e}=-1/4$ which
equals the lowest energy of the on-site potential, but is above the ground
state energy $v_{0}(\mu )$ and (ii) $\bar{e}(n_{s})$ is not the inverse of $%
\bar{n}_{s}(e)$. This difference is related to the fact that the maximum of $%
p(e,n_{s})$ with respect to $n_{s}$ for fixed $e$ is not generally related
to its maximum with respect to $e$ for fixed $n_{s}$.

\begin{figure}
\unitlength1cm
\begin{picture}(11,6)
\centerline{\psfig{file=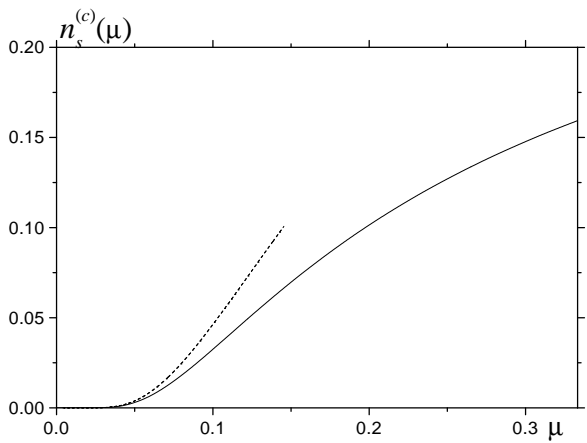,angle=-90,width=9cm}}
\end{picture}
\caption{\label{fig3} $\mu$-dependence of the critical average saddle index $n_{s}^{(c)} (\mu)$ for $0
\leq \mu \leq 1/3$ (solid line). The dashed line is the asymptotic
result Eq.~(\protect\ref{critParAnal}) for $\mu \ll 1$.}
\end{figure}%
%

\begin{figure}
\unitlength1cm
\begin{picture}(11,6)
\centerline{\psfig{file=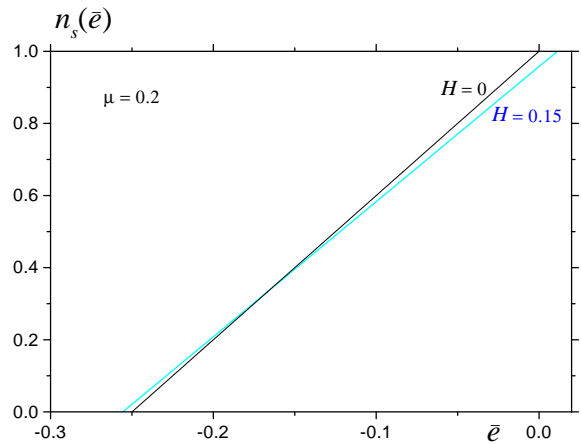,angle=-90,width=9cm}}
\end{picture}
\caption{\label{fig6} Saddle index $n_s(\bar e)$ versus
averaged energy $\bar e$ for $\mu=0.2$ and $H=0$ and 0.15.}
\end{figure}

\begin{figure}
\unitlength1cm
\begin{picture}(11,6)
\centerline{\psfig{file=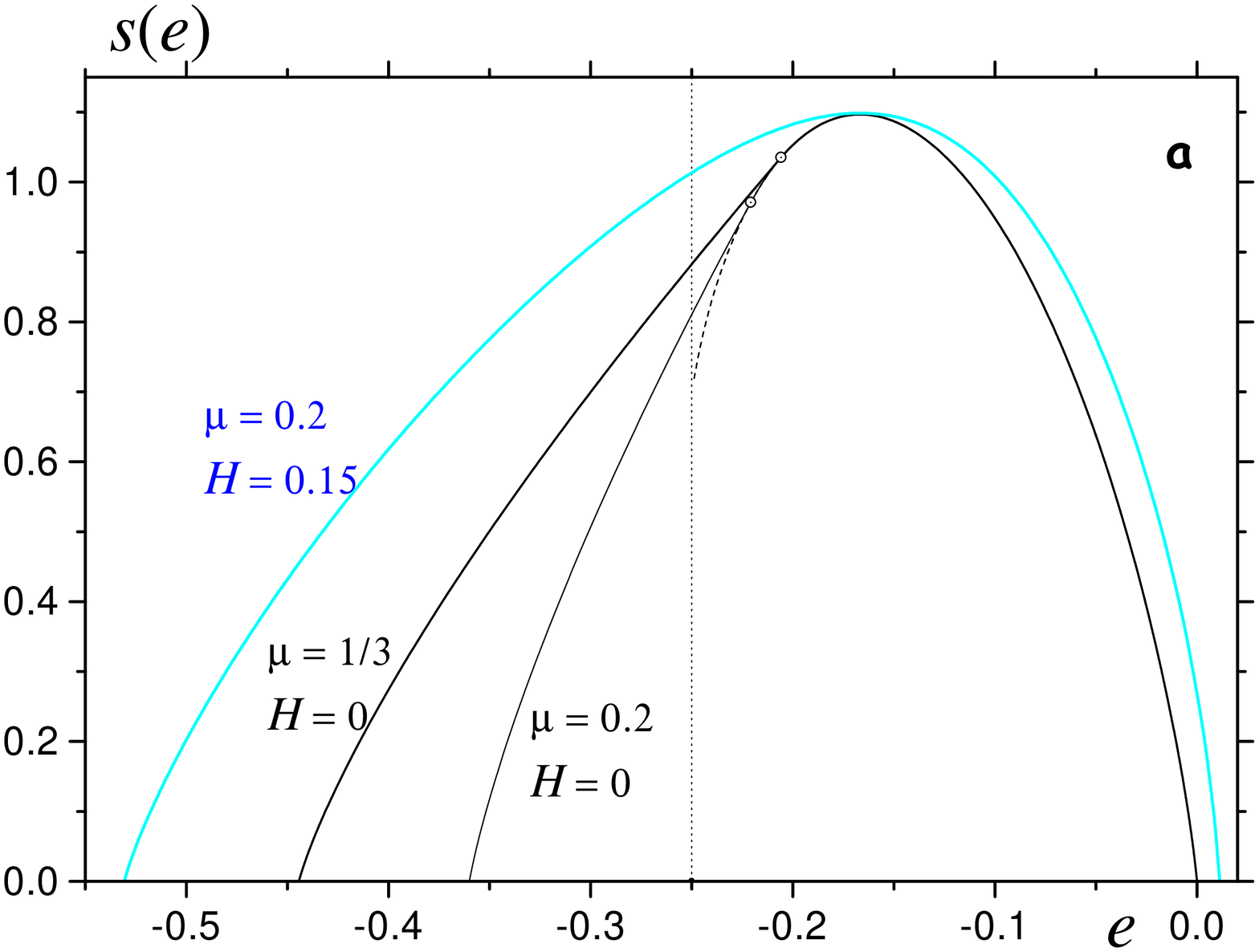,angle=-90,width=9cm}}
\end{picture}
\begin{picture}(11,6)
\centerline{\psfig{file=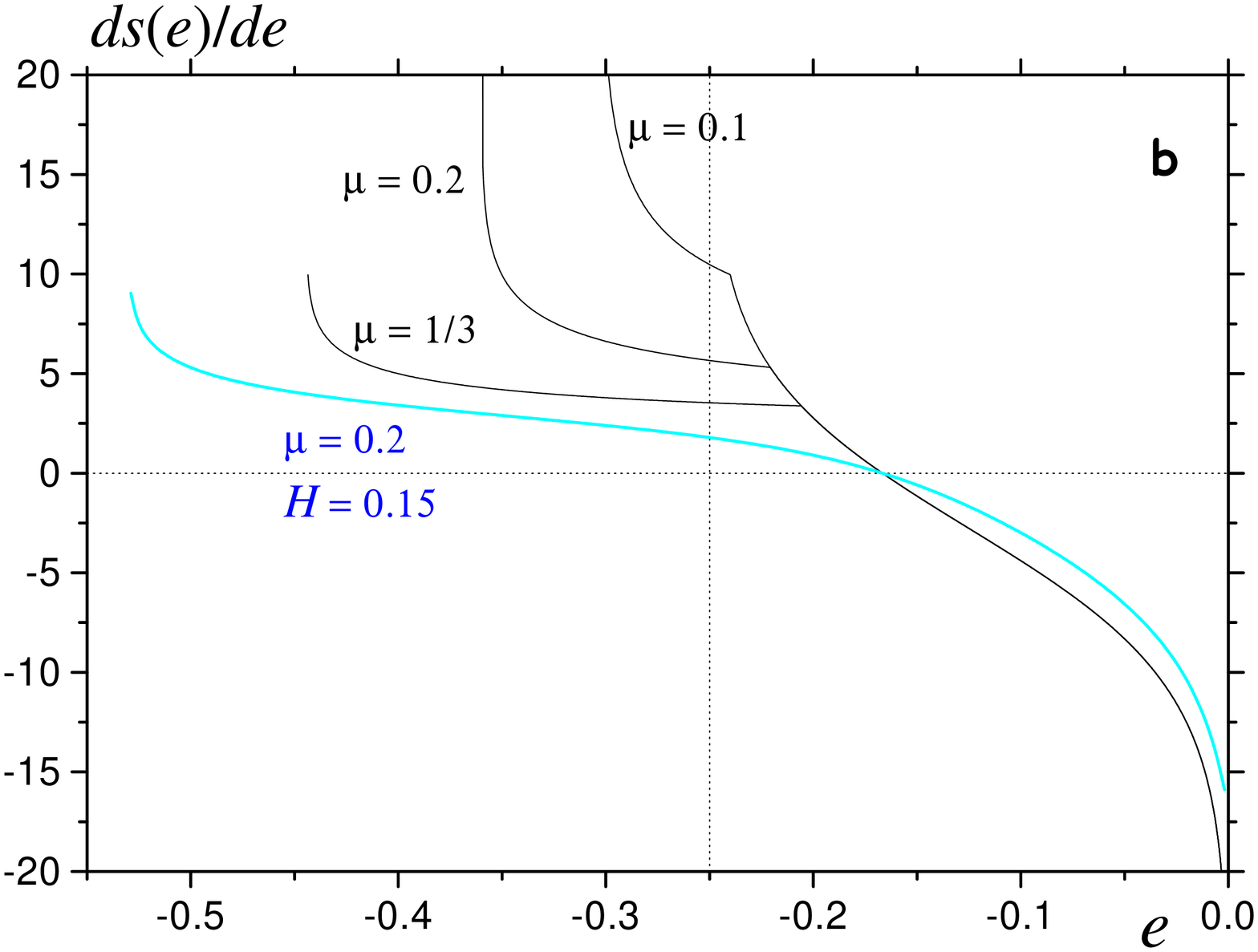,angle=-90,width=9cm}}
\end{picture}
\caption{\label{fig5} $a$ -- Energy dependence of the configurational entropy $s(e)$
for different interaction strength $\mu$, with and without field.
Dashed line is the solution for $\mu=0$ and $H=0$ and the circle indicates the location of
$e_c(\mu)$;
$b$ -- Derivative of the configurational entropy showing a transition
at $e=e_c(\mu)$ for $H=0$.
}
\end{figure}

Now let us take $H\neq 0$ (see Fig. \ref{fig4}b). As discussed in Appendix
B, $s(e,n_{s})$ has always a maximum at $\bar{n}_{s}(e)$ as function of $%
n_{s}$ within its domain. $\bar{n}_{s}(e)$ is shown in Fig. \ref{fig4}b. The
nonsingular $e$-dependence for $H\neq 0$ can clearly be seen. The average
energy $\bar{e}(n_{s})$ or its inverse $n_{s}(\bar{e})$ can be derived
analytically (see Appendix B). The result for $n_{s}(\bar{e})$ is shown in
Fig. \ref{fig6}.

Having determined $\bar{n}_{s}(e)$ for $H=0$ and $H\neq 0,$ one can now
calculate the energy-dependent configurational entropy that follows from Eq.
(\ref{pensDef}), in the limit $N\rightarrow \infty $
\begin{equation}
s(e)=s(e,\bar{n}_{s}(e)).  \label{sefromsens}
\end{equation}
The result is shown in Fig. \ref{fig5}. As can be seen from $\ $Fig. \ref
{fig5}$b$,\ $s(e)$ has a discontinuous second derivative for $H=0$ at $%
e=e_{c}(\mu ).$ In Fig. \ref{fig5}$a$ these points are marked by circles.
The high-energy branch of $s(e)$ has the form of Eq. (\ref{eq29}) with $\bar{%
n}_{s}(e)\Rightarrow 1+4e.$ It attains a maximum for $\bar{n}_{s}(e)=1/3$
that implies $e=-1/6,$ independently of the interaction.

\bigskip
%

%

\subsection{ Euler characteristic $\protect\chi (e)$}

It turns out that for the model under consideration in the limit $%
N\rightarrow \infty $ the Euler characteristic of Eq. (\ref{eq1}) satisfies
\begin{equation}
\left| \chi (e)\right| \sim p(e),\qquad \frac{1}{N}\ln \left| \chi
(e)\right| =s(e).  \label{chiRes}
\end{equation}
Calculation of $\chi (e)$ for large $N$ is similar to that of $s(e),$ as
suggested by the similar form of Eqs. (\ref{eq2}) and (\ref{eq3}). The only
difference is that $M(e,N_{s})$ used in the calculation of $\chi (e)$ is,
unlike $p(e,n_{s}),$ nonzero for $\upsilon _{\mathrm{max}}(\alpha _{0})<e.$
In this case it is independent of the energy and has the form
\begin{equation}
M(N_{s})=\frac{2^{N-N_{s}}N!}{N_{s}!(N-N_{s})!},  \label{MNsAbove}
\end{equation}
similar to Eq. (\ref{eq28}). Hence there are two different contributions
into $\chi (e).$ It can be easily shown that the contribution from the range
$\upsilon _{\min }(\alpha _{0})<e<\upsilon _{\mathrm{max}}(\alpha _{0})$ for
$N\rightarrow \infty $ coincides with that of $p(e)\sim \exp \left[ Ns(e)%
\right] $ that was studied above up to an irrelevant prefactor$,$ in spite
of the sign alternation in Eq. (\ref{eq1}). In contrast, the contribution
from the range of $N_{s}$ determined by $\upsilon _{\mathrm{max}}(\alpha
_{0})<e$ is dominated \emph{not} by the maximum of $M(N_{s})$ on $N_{s}$ but
by $N_{s}$ on the boundary of its interval, i.e., by $N_{s}$ satisfying $%
\upsilon _{\mathrm{max}}(N_{s}/N)=e.$ One can easily see that this
contribution to $\chi (e)$ never exceeds that from the range $\upsilon
_{\min }(\alpha _{0})<e<\upsilon _{\mathrm{max}}(\alpha _{0}),$ thus one
obtains $\left| \chi (e)\right| \sim p(e)$ for $N\rightarrow \infty .$

The reason for such a peculiar behavior of the contribution from the region $%
\upsilon _{\mathrm{max}}(\alpha _{0})<e$ is the sign alternation in $\chi
(e) $ plus the specific form of $M(N_{s}).$ For instance, as all stationary
points are below the level $e=0,$ one finds $\chi (e)$ for $e>0,$ by just
summing over all $N_{s}$:
\begin{equation*}
\chi (e>0)=\sum\limits_{N_{s}=0}^{N}(-1)^{N_{s}}M(N_{s})=1.
\end{equation*}
This result is exact and it has a transparent topological meaning. Replacing
the sum by the maximal summand value $M(N/3)$ would be an error. Even
simplifying Eq. (\ref{MNsAbove}) with the help of the Stirling formula for $%
N\gg 1$ in the sum would lead to an exponentially large result instead of 1.
Therefore, one should be cautious in applying the saddle point method to the
r.h.s. of Eq.~(\ref{eq1}).

\section{Discussion}

We have investigated the statistics of stationary points and topological
properties of the analytically tractable potential energy surface of a $\phi
^{4}$ model in a symmetry-breaking field $H$ with interaction of all pairs
of particles with the same strength $\mu $. For this model the mean-field
approximation becomes exact in the thermodynamic limit $N\rightarrow \infty $%
. For $H=0$ there is a second-order phase transition at the critical
temperature $T_{c}(\mu )$ that is analytical in $\mu $.

We have shown that the distribution of the saddle indices $p(n_{s}),$ where $%
n_{s}=N_{s}/N$ and $N_{s}$ is the number of unstable directions at a
stationary point has a maximum at $n_{s}=n_{s}^{\mathrm{max}}=1/3$.
Interestingly this value is consistent with that found for small binary
Lennard-Jones clusters \cite{17}. Whether or not this is an accident is not
clear. Our result originates from the fact that all stationary points can be
labelled by symbolic sequences $(\sigma _{1},\cdots ,\sigma _{n})$ with $%
\sigma _{n}=+,0,-$. The low-temperature anomalies of structural glasses are
usually explained by the existence of two-level systems arising from an
ensemble of double-well potentials. As the smallest ``unit'' of a PES of a
classical $N$-particle system, one may choose the local minima including
their basins of attraction. But such a choice does not fully encompass the
saddle. Taking the next larger unit, a pair of local minima and their common
saddle, one arrives at a double-well characterization of the PES. This could
explain why $n_{s}^{\max }=1/3$ for small clusters and for liquids. \ We
have also argued that the value $n_{s}^{\mathrm{max}}=1/3$ is a topological
invariant for an entire family of $\phi ^{4}$ models. In any case, it would
be important to determine $n_{s}^{\mathrm{max}}$ for other e.g., liquid-like
models and to check whether it equals again $n_{s}^{\mathrm{max}}=1/3$.

For our model the absolute value of the Euler characteristic $\chi (e)$ is
essentially the same as the density of stationary points $p(e)$ in the limit
$N\rightarrow \infty $, see Eq. (\ref{chiRes}). It would be interesting to
investigate the generality of this result.

For $H=0$ we have found a singularity in $p(e)$ and thus in $\left| \chi
(e)\right| $ at the energy $e_{c}(\mu )$ given by Eq. (\ref{critParAnal}). $%
e_{c}(\mu )$ is nonanalytic in $\mu .$ At $e=e_{c}(\mu )$ the second
derivative $d^{2}\ln \left| \chi \right| (e)/de^{2}$ is discontinuous, as
found for the models studied in Refs. \cite{10,11,12,13}. In agreement with
these papers, we also have found that the topological singularity disappears
for nonzero field, as well as the thermodynamic singularity. In this
respect, we would like to mention a recent publication Ref. \cite{26}, where
it is proven that a topological singularity is a \textit{necessary}
condition for a thermodynamical transition to take place. However, at
variance with earlier work \cite{10,11,12,13}, this singularity is not
related to the thermodynamic singularity of our model as $e_{c}(\mu )$ does
not coincide with $\upsilon _{c}(\mu ),$ the average potential energy at
temperarure $T_{c}(\mu ).$

The reasons for this discrepancy can be made clear with the help of the
following argument. Let us smoothly modify the local potential $V_{0}(x)$ in
the intervals $[-\infty ,-1-\varepsilon (\mu )]$, $[-1+\varepsilon (\mu
),-\varepsilon (\mu )]$, $[\varepsilon (\mu ),1-\varepsilon (\mu )]$ and $%
[1+\varepsilon (\mu ),\infty ]$ for given $0<\varepsilon (\mu )<1$. If $\mu $
is small enough, which implies that the internal field is small enough, then
the three roots of Eq.~(\ref{eq13}) are within the intervals $[\sigma
-\varepsilon (\mu ),\sigma +\varepsilon (\mu )]$, $\sigma =+,0,-_{,}$ in
which the potential has not been modified. Accordingly, the roots and
therefore the stationary points and their energies are the same. This
implies that $e_{c}(\mu )$ is the same. However, since the calculation of $%
T_{c}(\mu )$ [cf.~Eq.~(\ref{eq9})] involves $V_{0}(x)$ for \textit{all} $x$,
the critical temperature will be different for the modified on-site
potential.

The discrepancy between the topological and thermodynamic singularities can
also be traced back to an unjustified comparison of a continuous model
(thermodynamics) and a discrete model (topology). More logically, the energy
of all the stationary points can be represented by an Ising-like Hamiltonian
$\mathcal{H}(\{\sigma _{i}\})$ (for $\mu <1/3)$ with $\sigma _{i}=+,0,-.$
The corresponding canonical partition function $\mathcal{Z}(T)=\mathrm{Tr}%
\exp \left[ -\mathcal{H}(\{\sigma _{i}\})\right] $ can be calculated from
the density of states $p(e)$ as
\begin{equation}
\mathcal{Z}(T)=\int dep(e)\exp \left[ -Ne/T\right] .  \label{ZTviape}
\end{equation}
Evidently a singularity of $\mathcal{Z}(T)$ at the corresponding transition
temperature $T_{c}^{\prime }$ results from the underlying singularity of $%
p(e)$ at $e_{c}.$ Obviously in this case $\upsilon _{c}^{\prime }(\mu
)=\langle \mathcal{H}\rangle (T_{c}^{\prime })/N=e_{c}(\mu )$ is fulfilled.
But the thermodynamics of this discrete model does not coincide with that of
the original continuous model, in particular, $T_{c}^{\prime }\neq T_{c}$.

The idea of an at least qualitative relationship between the thermodynamic
singularity and the topological singularity is supported by the following
observation. At the thermodynamic transition point there appears a
spontaneous breaking of the left-right symmetry for the displacements, which
is equivalent to emerging of a nonzero temperature-dependent internal field
for $T<T_{c}(\mu ).$ On the other hand, the stationary configurations with $%
e>e_{c}(\mu )$ and with maximum weight correspond to $\alpha _{+}=$ $\alpha
_{-}.$ This implies that the effective field defined by Eqs. (\ref{eq12})
and (\ref{eq14}) satisfies $H_{\mathrm{eff}}=0.$ However for $e<e_{c}(\mu )$
it is $\alpha _{+}\neq $ $\alpha _{-}$ and hence $H_{\mathrm{eff}}\neq 0.$
Therefore a spontaneous symmetry breaking occurs at both singularities.

The energy- or temperature-dependence of the averaged saddle index $\bar{n}%
_{s}$ seems to play a role for the dynamical features of supercooled
liquids. For the present model we have found that $\bar{n}_{s}(e)$ vanishes
at the ground state energy $v_{0}$, only. This is consistent with recent
results on $\bar{n}_{s}(T)$ showing that $\bar{n}_{s}=0$ at $T=0$, only \cite
{13}. Taking the analogy to mean-field like spin glass models, this would
imply that no dynamical transition (or crossover) could occur at finite
temperatures\cite{CGP}. The averaged energy $\bar{e}(n_{s})$ as function of $%
n_{s}$ is \emph{not} the inverse function of $\bar{n}_{s}(e)$. In particular
$n_{s}(\bar{e})$ vanishes at $e^{\ast }(H)>v_{0}$. Whether dynamics changes
qualitatively at $e^{\ast }(H)$ or any other characteristic temperature
would be interesting to study.

Finally we would like to mention that after submission of this paper we
learned about a similar study of the same model \cite{Ruocco}, where the
authors find, among others, the same type of discrepancy between the
thermodynamical and topological singularities.

\appendix

\section{Properties of $v(\protect\alpha _{+},\protect\alpha _{0})$}

In this Appendix we investigate the properties of $v(\alpha _{+},\alpha
_{0}) $ defined by Eq. (\ref{vgenaral}). For the discussions in Sec. 3, the
first and second derivative of $\upsilon (\alpha _{+},\alpha _{0})$ with
respect to $\alpha _{+}$ will be useful. A straightforward calculation
making use of Eqs.~(\ref{eq13}),~(\ref{eq14}) and~(\ref{eq18}) yields:

\begin{equation}
\frac{\partial \upsilon (\alpha _{+},\alpha _{0})}{\partial \alpha _{+}}%
=V_{0}(x_{+}(H_{\mathrm{eff}}),H_{\mathrm{eff}})-V_{0}(x_{+}(H_{\mathrm{eff}%
}),H_{\mathrm{eff}}).
\end{equation}
Since $H_{\mathrm{eff}}=0$ implies $x_{\pm }=\pm 1$ [see Eq. (\ref{eq13})]
it follows that $\partial \upsilon (\alpha _{+},\alpha _{0})/\partial \alpha
_{+}=0$ for $H_{\mathrm{eff}}=0$. That is, the maximum of $\upsilon (\alpha
_{+},\alpha _{0})$ with respect to $\alpha _{+}$ corresponds to $H_{\mathrm{%
eff}}=0$. For the second derivative we get:
\begin{eqnarray}
\frac{\partial ^{2}\upsilon (\alpha _{+},\alpha _{0})}{\partial \alpha
_{+}^{2}} &=&-[x_{+}(H_{\mathrm{eff}})-x_{-}(H_{\mathrm{eff}})]  \notag \\
&&\qquad \times \frac{\partial H_{\mathrm{eff}}(\alpha _{+},\alpha _{0})}{%
\partial \alpha _{+}}  \label{eq24}
\end{eqnarray}
with
\begin{equation}
\frac{\partial H_{\mathrm{eff}}(\alpha _{+},\alpha _{0})}{\partial \alpha
_{+}}=\mu \frac{x_{+}(H_{\mathrm{eff}})-x_{-}(H_{\mathrm{eff}})}{1-\mu
\sum\limits_{\sigma }\frac{\alpha _{\sigma }}{3x_{\sigma }^{2}(H_{\mathrm{eff%
}})-1}}.  \label{eq25}
\end{equation}
Since $x_{+}-x_{-}>0$, the ``curvature'' $\partial ^{2}\upsilon /\partial
\alpha _{+}^{2}$ is negative, i.e.,~$\upsilon (\alpha _{+},\alpha _{0})$ is
concave in $\alpha _{+}$, provided $\partial H_{\mathrm{eff}}/\partial
\alpha _{+}>0$. This is true if $\mu $ is small enough, as can be seen from
Eq.~(\ref{eq25}).

The function $\upsilon (\alpha _{+},\alpha _{0})$ can be computed
analytically near its maximum in $\alpha _{+}$ by using the expansions
\begin{eqnarray}
x_{-}(H_{\mathrm{eff}}) &\cong &-1+H_{\mathrm{eff}}/2  \notag \\
x_{0}(H_{\mathrm{eff}}) &\cong &-H_{\mathrm{eff}}  \notag \\
x_{+}(H_{\mathrm{eff}}) &\cong &1+H_{\mathrm{eff}}/2  \label{xHSolTailor}
\end{eqnarray}
near $H_{\mathrm{eff}}=0.$ The result is a parabola in $\alpha _{+}$:
\begin{eqnarray}
\tilde{\upsilon}(\alpha _{+},\alpha _{0}) &\cong &-\frac{1}{4}\left(
1-\alpha _{0}\right) +\frac{H^{2}}{2\mu }  \notag \\
&&-\frac{1}{2\mu }\frac{\left[ H+\mu (2\alpha _{+}+\alpha _{0}-1)\right] ^{2}%
}{1-\mu (1-3\alpha _{0})/2}.  \label{EnergyParabola}
\end{eqnarray}
Analysis shows that corrections to this formula in the whole region $0\leq
\alpha _{+}\leq 1-\alpha _{0}$ are of order $\mu ^{3},$ $\mu ^{2}H,$ and $%
\mu H^{2},$ i.e., Eq. (\ref{EnergyParabola}) is a very good approximation
for not too large $\mu .$ For instance, the ground-state energy following
from Eq. (\ref{EnergyParabola}) in the case $H=0$%
\begin{equation*}
\tilde{\upsilon}_{0}(\mu )=\tilde{\upsilon}(0,0)=-\frac{1}{4}\left( 1+\frac{%
2\mu }{1-\mu /2}\right)
\end{equation*}
is in accord with the exact two-fold degenerate ground-state energy,
\begin{equation}
\upsilon _{0}(\mu )=\upsilon (1,0)=\upsilon (0,0)=-\frac{1}{4}(1+\mu )^{2}.
\label{eq26}
\end{equation}
up to the terms $\mu ^{2},$ and the relative error is only $0.0125$ for $\mu
=1/3.$ The exact value of the field-dependent maximal potential energy that
also follows from Eq. (\ref{EnergyParabola}) has the form
\begin{equation}
\upsilon _{\mathrm{max}}(\alpha _{0})=-\frac{1}{4}\left( 1-\alpha
_{0}\right) +\frac{H^{2}}{2\mu }.  \label{eq27}
\end{equation}
The corresponding exact value of $\alpha _{+}$ is
\begin{equation}
\alpha _{+}^{(\max )}(\alpha _{0})=\frac{1-\alpha _{0}}{2}-\frac{H}{2\mu }.
\label{alphaPlusMax}
\end{equation}
Eqs. (\ref{eq27}) and (\ref{alphaPlusMax}) are thus valid for $|H|\leq \mu
(1-\alpha _{0}).$

\section{Configurational entropy}

In this Appendix we present analytical results for the configurational
entropy $s(e,n_{s}).$ In particular, for $H=0$ the existence of a critical
energy $e_{c}(\mu )$ will be proven, at which singular energy dependence
occurs.

First we have to investigate $\upsilon (\alpha _{+},\alpha _{0}).$ It is
convenient to use the approximate form of $\upsilon (\alpha _{+},\alpha _{0})
$ given by Eq. (\ref{EnergyParabola}). Then Eq. (\ref{eq32}) becomes
quadratic, and one obtains two solutions
\begin{eqnarray}
\alpha _{+}^{(\pm )} &=&\alpha _{+}^{(\max )}(\alpha _{0})  \notag \\
&&\pm \sqrt{\frac{\left[ \upsilon _{\max }\left( \alpha _{0}\right) -e\right]
\left[ 1-\mu (1-3\alpha _{0})/2\right] }{2\mu }},  \label{alphaplusEAppr}
\end{eqnarray}
where $\alpha _{+}^{(\max )}(\alpha _{0})$ and $\upsilon _{\max }\left(
\alpha _{0}\right) $ are given by Eqs. (\ref{alphaPlusMax}) and (\ref{eq27}%
), respectively. Note that this result becomes exact for all $\mu $ with $%
0<\mu \leq 1/3$ provided that $e$ is close to $\upsilon _{\mathrm{max}%
}(\alpha _{0})$.

The density of stationary points $p(e,n_{s})$ can be used to calculate the
statistics of saddle indices, see Eqs. (\ref{nsAvrDef})--(\ref{smaxEqs}).
The value of $\bar{n}_{s}(e)$ that maximizes the entropy $s\left( e,\alpha
_{0}\right) $ on $\alpha _{0}$ is the solution of the equation
\begin{equation}
\frac{\partial \alpha _{+}^{(+)}}{\partial \alpha _{0}}\ln \frac{1-\alpha
_{0}-\alpha _{+}^{(+)}}{\alpha _{+}^{(+)}}+\ln \frac{1-\alpha _{0}-\alpha
_{+}^{(+)}}{\alpha _{0}}=0  \label{Deralpha0Eq}
\end{equation}
for $\alpha _{0},$ if this solution exists in the interval of Eq. (\ref
{alpha0window}). Otherwise the maximum of $s\left( e,\alpha _{0}\right) $ is
attained at the left boundary of the $\alpha _{0}$-window, $\alpha
_{0}=\alpha _{0}^{(\min )}(e)$ (see Fig. \ref{fig2}$a)$. The latter solution
exists only for $H=0.$ It corresponds to the maxima of the curves $\upsilon
(\alpha _{+},\alpha _{0})$ in Fig. \ref{fig1}$a,$ i.e., to
\begin{equation}
\alpha _{+}^{(+)}=\alpha _{+}^{(-)}=\alpha _{+}^{(\max )}(\alpha _{0})=\frac{%
1-\alpha _{0}}{2}.  \label{maxsolution}
\end{equation}
One can see from Eq. (\ref{alphaplusEAppr}) that $\partial \alpha
_{+}^{(+)}/\partial \alpha _{0}$ diverges for $e\rightarrow \upsilon _{\max
}\left( \alpha _{0}\right) ,$ i.e., for $\alpha _{0}\rightarrow \alpha
_{0}^{(\min )}(e).$ One can check that in the case $H=0$ this divergence is
compensated for by the log factor that tends to zero. For $H\neq 0$ there is
no such compensation, and the lhs of Eq. (\ref{Deralpha0Eq})\ diverges at $%
\alpha _{0}\rightarrow \alpha _{0}^{(\min )}(e),$ thus Eq. (\ref{Deralpha0Eq}%
) always has a solution, see Fig. \ref{fig2}$b.$

Let us consider the case $H=0$ and find the condition that the solution of
Eq. (\ref{Deralpha0Eq}) is just $\alpha _{0}=\alpha _{0}^{(\min )}(e).$
Simplifying this equation for $e\rightarrow \upsilon _{\max }\left( \alpha
_{0}\right) $ one obtains the transcedental equation
\begin{equation}
-\frac{1-\mu (1-3\alpha _{0})/2}{4\mu \left( 1-\alpha _{0}\right) }+\ln
\frac{1-\alpha _{0}}{2\alpha _{0}}=0.  \label{alpha0muEq}
\end{equation}
Its solution $\alpha _{0}^{(c)}(\mu )\equiv n_{s}^{(c)}(\mu )$ that is
plotted in Fig. \ref{fig3} and the corresponding energy $e^{(c)}(\mu )$ are
critical parameters that define the boundary between different regimes. In
particular, $\alpha _{0}^{(c)}(1/3)\simeq 0.1774$ and $e^{(c)}(1/3)\simeq
-0.2056.$ For $\mu \ll 1$ one can solve Eq. (\ref{alpha0muEq}) analytically:
\begin{gather}
n_{s}^{(c)}(\mu )\equiv \alpha _{0}^{(c)}(\mu )\cong \frac{1}{2}\exp \left( -%
\frac{1}{4\mu }\right)  \notag \\
e_{c}(\mu )=-\frac{1-n_{s}^{(c)}(\mu )}{4}\cong -\frac{1}{4}+\frac{1}{8}\exp
\left( -\frac{1}{4\mu }\right) .  \label{critParAnal}
\end{gather}
Note that $e^{(c)}(\mu )$ is always above $-1/4,$ the ground-state energy
without interaction. Now one can write down the combined expression for $%
\bar{n}_{s}(e)$ in the case $H=0$:
\begin{equation}
\bar{n}_{s}(e)=\left\{
\begin{array}{cc}
\bar{n}_{s}^{<}(e), & \upsilon _{0}(\mu )\leq e\leq e_{c}(\mu ) \\
\bar{n}_{s}^{\max }(e), & e_{c}(\mu )\leq e\leq 0 \\
0, & 0\leq e.
\end{array}
\right.  \label{nsavrResFinaö}
\end{equation}
Here $\upsilon _{0}(\mu )$ is the ground-state energy, $\bar{n}_{s}^{<}(e)$
is equal to $\alpha _{0}$ that solves Eq. (\ref{Deralpha0Eq}), and
\begin{equation}
\bar{n}_{s}^{\max }(e)=1+4e.  \label{nsmax}
\end{equation}
Note that the high-energy branch $\bar{n}_{s}^{\max }(e)$ is independent of $%
\mu $ and is thus the same as for a system of noninteracting particles. This
contribution is due to the maxima of $\upsilon (\alpha _{+},\alpha _{0})$
[see Eq. (\ref{maxsolution})] whereas $\bar{n}_{s}^{<}(e)$ is the
contribution from the energy levels below this maximum. One can show that $%
\bar{n}_{s}(e)\rightarrow 0$ for $e\rightarrow \upsilon _{0}(\mu ).$ This
dependence has the form given by Eq. (\ref{34}). However this dependence is
only realized for the energies very close to the ground state $\upsilon
_{0}(\mu )$. The function $\bar{n}_{s}(e)$ is shown in Fig. \ref{fig4} for $%
H=0$ and $H\neq 0.$ It has a discontinuous derivative at $e=e_{c}(\mu )$ in
zero field. This discontinuity disappears for $H\neq 0.$

One can also use $p(e,n_{s})$ to calculate the average energy $\bar{e}%
(n_{s}) $ for a given $n_{s}.$ From the second of Eqs. (\ref{smaxEqs}) one
obtains
\begin{equation*}
\ln \frac{1-\alpha _{0}-\alpha _{+}^{(+)}}{\alpha _{+}^{(+)}}=0
\end{equation*}
that is simpler than Eq. (\ref{Deralpha0Eq}). From its solution
\begin{equation}
\bar{e}(n_{s})=-\frac{1-n_{s}}{4}-\frac{H^{2}}{4}\frac{1-3n_{s}}{1-\mu
(1-3n_{s})/2}  \label{eavrvsns}
\end{equation}
one obtains $n_{s}(\bar{e})$ shown in Fig. \ref{fig6} that is almost linear
in $\bar{e}$ in the considered range of parameters, Eq. (\ref{eq22}). Note
that, in contrast to $\bar{n}_{s}(e),$ the quantity $n_{s}(\bar{e})$ found
from Eq. (\ref{eavrvsns}) turns to zero not at the ground-state energy but
at $\bar{e}=-1/4-H^{2}/\left[ 2(2-\mu )\right] $.


\begin{thebibliography}{99}
\bibitem{24}  Ref. \cite{23} has studied $H=0$, only. But it is
straightforward to generalize those results to $H\neq 0$.
\bibitem{25}  That the stationary points for $H=0$ are uniquley determined
by $\{\alpha _{\sigma }\}$ has been rigorously proven in Ref. \cite{23} for $%
\alpha _{0}=0$, only. But the numerical solution of Eq.~(\ref{eq18}) for $%
\alpha _{0}\neq 0$ and $H\neq 0$ has always led to a unique solution $H_{%
\mathrm{eff}}(\alpha _{+},\alpha _{0},H)$. Therefore we will assume that $%
\{\alpha _{\sigma}\}$ is in a one-to-one correspondence with all stationary
states as long as Eq.~(\ref{eq22}) is fulfilled.
\end{thebibliography}
\end{document}